# Controllability analysis of the directed human protein interaction network identifies disease genes and drug targets

Arunachalam Vinayagam[1,*], Travis E. Gibson[2], Ho-Joon Lee[3], Bahar Yilmazel[4,5], Charles Roesel[4,5,†], Yanhui Hu[1,4], Young Kwon[1], Amitabh Sharma[2,6,7], Yang-Yu Liu[2,6,7,*], Norbert Perrimon[1,8*] and Albert-László Barabási[6,7*]

[1]Department of Genetics, Harvard Medical School, 77 Avenue Louis Pasteur, Boston, MA 02115, USA

[2]Channing Division of Network Medicine, Brigham and Women's Hospital, Harvard Medical School, Boston, MA 02115, USA

[3]Department of Systems Biology, Harvard Medical School, 200 Longwood Avenue, Boston, MA 02115, USA

[4]Drosophila RNAi Screening Center, Department of Genetics, Harvard Medical School, 77 Avenue Louis Pasteur, Boston, MA 02115, USA

[5]Bioinformatics program, Northeastern University, 360 Huntington Avenue, Boston, MA 02115, USA

[6]Center for Complex Network Research and Department of Physics, Northeastern University, Boston, MA , USA

[7]Center for Cancer Systems Biology, Dana-Farber Cancer Institute, Boston, MA, USA

[8]Howard Hughes Medical Institute, 77 Avenue Louis Pasteur, Boston, MA 02115, USA

*Correspondence to: Arunachalam Vinayagam (vinu@genetics.med.harvard.edu), Yang-Yu Liu (yyl@channing.harvard.edu), Norbert Perrimon (perrimon@receptor.med.harvard.edu), Albert-László Barabási (alb@neu.edu)

† Present address: Marine Science Center, Northeastern University, Nahant, MA 01908, USA

**One Sentence Summary:** Controllability analysis of large-scale protein interaction network identifies disease genes.

**Abstract**

The protein-protein interaction (PPI) network is crucial for cellular information processing and decision-making. With suitable inputs, PPI networks drive the cells to diverse functional outcomes such as cell proliferation or cell death. Here we characterize the structural controllability of a large directed human PPI network comprised of 6,339 proteins and 34,813 interactions. This allows us to classify proteins as "indispensable", "neutral" or "dispensable", which correlates to increasing, no effect, or decreasing the number of driver nodes in the network upon removal of that protein. We find that 21% of the proteins in the PPI network are indispensable. Interestingly, these indispensable proteins are the primary targets of disease-causing mutations, human viruses, and drugs, suggesting that altering a network's control property is critical for the transition between healthy and disease states. Furthermore, analyzing copy number alterations data from 1,547 cancer patients reveals that 56 genes that are frequently amplified or deleted in nine different cancers are indispensable. Among the 56 genes, 46 of them have not been previously associated with cancer. This suggests that controllability analysis is very useful in identifying novel disease genes and potential drug targets.

**Significance statement**

Large-scale biological network analyses often employ concepts used in social networks analysis, e.g. finding "communities", "hubs", etc. However, mathematically advanced engineering concepts have only been applied to analyze small and well-characterized networks so far in biology. Here, we applied a sophisticated engineering tool, from control theory, to analyze a large-scale directed human protein-protein interaction network. Our analysis revealed that the proteins that are indispensable, from a network controllability perspective, are also commonly targeted by disease causing mutations, human viruses, or have been identified as drug targets. Furthermore, we used the controllability analysis to prioritize novel cancer genes that were mined from cancer genomic datasets. Altogether we demonstrated a novel application of network controllability analysis to identify new disease genes and drug targets.

**Introduction**

The need to control engineered systems has resulted in a mathematically rich set of tools that are widely applied in the design of electric circuits, manufacturing processes, communication systems, aircraft, spacecraft, and robots(1-3). Control theory deals with the design and stability analysis of dynamic systems that receive information via inputs and have outputs available for measurement. Issues of control and regulation are central to the study of biological systems(4, 5), which sense and process both external and internal cues using a network of interacting molecules(6). The dynamic regulation of this molecular network in turn drives the system to various functional states - such as triggering cell proliferation or inducing apoptosis. This feature of specific input signals driving networks from an initial state to a specific functional state suggests that the need to control a biological system plays a potentially important role in the evolution of molecular interaction networks. Note that the term *state* is also used in a control context where the *state space* of a control system is the space of values the state variables can attain. For a protein-protein interaction (PPI) network, the *state variables* are the specific protein concentrations and the state space is all positive real numbers of dimension equal to the total number of proteins in the PPI network.

According to control theory, a dynamic system is controllable if, with a suitable choice of inputs, it can be driven from any initial state to any desired final state in finite time(2, 7). Previous studies have shown that network components exhibit properties of control systems such as proportional action, feedback control, and feed-forward control(8-12). However the main challenges that hinder systematic controllability analysis of biological networks are the availability of large-scale biologically relevant networks and efficient tools to analyze their controllability. To address these issues, two resources were integrated in this work: (i) a directed human PPI network(13); and (ii) an analytical framework to characterize the structural controllability of directed weighted networks(14). The directed human PPI network represents a global snapshot of the information flow in cell signaling. For a given weighted and directed network associated with linear time-invariant dynamics, the analytical framework identifies a minimum set of driver nodes, whose control is sufficient to fully control the dynamics of the whole network(14, 15).

In this work we classified the proteins (nodes) as *indispensable*, *neutral* or *dispensable*, based on the change of the minimum number of driver nodes needed to control the PPI network when a specific protein (node) is absent. In addition, we analyzed the role of different node types in the context of human diseases. Using known examples of disease causing mutations, virus-targets and drug-targets, we identified indispensable nodes that are key players in mediating the transition between healthy and disease states. Our study illustrates the potential application of network controllability analysis as a powerful tool to identify new disease genes.

## Results
### Characterizing the controllability of the directed PPI network

We applied linear control tools to access local controllability of PPI networks whose dynamics are inherently nonlinear. The experimentally obtained network, however, can be assumed to capture linear affects around homeostasis. Furthermore, given that the tools developed in (14) are for linear dynamics we are careful to only assume that we can ascertain local controllability around homeostasis. Controllability henceforth referred to local controllability (see Supplementary Text for details).

The directed human PPI network consists of 6,339 proteins (nodes) and 34,813 directed edges, where the edge direction corresponds to the hierarchy of signal-flow between the interacting proteins and the edge weight corresponds to the confidence of the predicted direction. We applied structural controllability theory to identify a minimum set of driver nodes - i.e., nodes through which we can achieve control of the whole network. Note that the identified minimum driver node set (MDS) is not unique, but its size, denoted as $N_D$, is uniquely determined by the network topology. We found that the MDS of the directed human PPI network contains 36% nodes. We also classified the nodes as *indispensable*, *neutral* or *dispensable*, based on the change of $N_D$ upon their removal. A node is (i) *indispensable* if removing it increases $N_D$ (e.g. node 2 in **Fig. 1a**); (ii) *neutral* if its removal has no effect on $N_D$ (e.g. node 1 in **Fig. 1a**); and (iii) *dispensable* if its removal reduces $N_D$ (e.g. nodes 3 and 4 in **Fig. 1a**). In the directed human PPI network, 21% of nodes are indispensable, 42% neutral, and the remaining 37% dispensable (**Fig. 1b**). Interestingly, we found that all the three node types have a heterogeneous degree distribution, and indispensable nodes tend to have higher in- and out-degrees compared to neutral and dispensable nodes (**Fig. 1b and 1c**). Similarly, indispensable nodes are associated with more PubMed records (http://www.ncbi.nlm.nih.gov/pubmed) and Gene Ontology(17) term annotation than neutral and dispensable nodes (**Fig. S1a-b**). However, the correlation between the node-degree and the literature bias is weak (correlation coefficient of 0.37 and 0.41 for in- and out-degree, respectively), suggesting that the higher degree of indispensable nodes is not explained by the literature bias alone (**Fig. S1c-d**).

We characterized indispensable, neutral and dispensable nodes in the context of essentiality, evolutionary conservation, and regulation at the level of translational and posttranslational modifications. Our gene essentiality analysis indicated that indispensable nodes are enriched in essential genes while essential genes are underrepresented among dispensable nodes (**Fig. 1e, Fig. S1e** and **Table S1**). Further, indispensable nodes are evolutionarily conserved from human to yeast compared to the other two node types (**Fig. 1e** and **Fig. S1f**). Next, we analyzed the different node types in the context of cell signaling, which is at the core of cellular information processing. In general, known signaling proteins are enriched as indispensable nodes. However, dissecting different functional classes within signaling proteins reveals that kinases are enriched as indispensable nodes whereas membrane receptors and transcription factors are enriched as neutral nodes (**Fig. 1e** and **Fig. S2a**). Analysis of the protein steady-state abundance in cell lines, as a measure of translational regulation, reveals that indispensable nodes are enriched as high copy number proteins whereas low copy number proteins show moderate enrichments for both indispensable and dispensable nodes (**Fig. 1e** and **Fig. S2b**). Similarly, indispensable nodes are highly regulated through posttranslational modification, including acetylation, ubiquitination and phosphorylation (pS/pT and pY) (**Fig. 1e** and **Fig. S2c**). Altogether, our enrichment analyses revealed distinct functional and regulatory roles for indispensable, neutral and dispensable nodes.

## Understanding healthy to disease state transition using network controllability

We analyzed the node classification in the context of driving the system from healthy to disease condition and vice-versa. Specifically, we analyzed the impact of three different transitions, 1) healthy to disease transition induced by mutations or other genetic alterations; 2) healthy to infectious transition induced by human viruses; and 3) disease to healthy transition induced by drugs or small molecules. Mutations or other genetic alterations were treated as external inputs to certain nodes in the network to drive the network from a healthy to a disease state. Our goal is to determine whether specific node types (indispensable, neutral or dispensable) are enriched for disease causing mutations. Our analysis of 445 genes annotated by the *Sanger Center* as causally implicated in oncogenesis (Cancer Gene Census; http://cancer.sanger.ac.uk/cancergenome/projects/census/)(18) revealed that indispensable nodes are highly enriched in cancer genes, whereas neutral nodes showed no enrichment and dispensable nodes are underrepresented (**Fig. 2a: Cancer I, Fig. S3a** and **Table S2**). To ensure that the observed enrichment of indispensable nodes is not due to the literature and degree bias, we repeated our analysis using literature and degree controlled random sets (see Methods). After adjusting for literature and degree bias (**Fig. 2a: PubMed, Degree** and **Table S2**), indispensable nodes remain significantly enriched as cancer genes. Note that for enrichment analysis below, the degree and literature controlled enrichments results were shown in Supplementary figures (**Fig. S3b**). To further substantiate that indispensable nodes are enriched as cancer genes, we analyzed 3,164 genes predicted as cancer related genes(19) and observed a similar enrichment for indispensable nodes (**Fig. 2b: Cancer II, Fig. S3a**).

Next, we analyzed 1,403 genes annotated by *Online Mendelian Inheritance in Man* (OMIM; http://omim.org/) as causal genes for various genetic diseases, aiming to test whether the perturbation of indispensable nodes is a specific feature of cancer or a general feature of human diseases. Our analysis showed that the perturbation of indispensable nodes is a common feature of human diseases (**Fig. 2b: OMIM, Fig. S3a**). Interestingly, however, our analysis of disease genes identified from genome-wide association studies (GWAS; www.genome.gov/gwastudies)(20) revealed poor enrichment for indispensable nodes (**Fig. 2b: GWAS, Fig. S3a**), most likely reflecting the fact that GWAS identify genomic regions but not specific coding genes that cause the disease(21). Since indispensable nodes are enriched for causal mutations (**Fig. 2a-b**), our resource could help identify causal genes from GWAS.

We also characterized the network controllability in the context of host-parasite interactions, specifically human-virus interactions. Upon infection, viruses control the host cellular network to utilize the host resources to replicate and to evade the host immune response. Here, we analyzed the node types targeted by human viruses to drive the network from a healthy state to an infectious state. First, we analyzed the targets of Human Immunodeficiency Virus (HIV), a member of the lentivirus family that causes Acquired Immunodeficiency Syndrome (AIDS). Putative human genes, identified to have an effect on HIV-1 replication from large-scale functional genomic screens (data compiled from 4 RNAi datasets)(22-25), tend to be indispensable nodes (**Fig. 2c: RNAi, Fig. S3c**). However, we did not detect a significant enrichment – most likely reflecting the quality of the HIV RNAi screens(26). To analyze direct targets of HIV, we compiled the HIV-human interactome (from recent literature and PPI databases)(27, 28), finding that indispensable nodes are enriched for physical interactions with HIV proteins (**Fig. 2c: PPIs, Fig. S3c**). Analysis of 208 different human-virus networks(27-30) reveals that human viruses commonly target indispensable nodes to control the host network (**Fig. 2c: Virus targets, Fig. S3c**). We noticed that after adjusting for literature bias indispensable nodes remain as viral targets, while adjusting for degree bias shows only weak enrichment (**Fig. S3d**). This is in agreement with the previous observations that viruses tend to target hubs(31).

Finally, we characterized the network controllability in the context of driving the system from disease to healthy state. Specifically, we analyzed the node types that are targeted by the drugs/small molecules (**Fig. 2d**). By analyzing the targets of drugs approved by Food and Drug Administration (FDA)(32), we found that indispensable nodes are enriched for drug targets (**Fig. 2d: FDA targets, Fig. S3e-f**). Extending the analysis to the list of proteins that are annotated as druggable(33), presence of protein folds that favor interactions with drug-like chemical compounds, showed that the druggable genome list is not significantly enriched for indispensable nodes (**Fig. 2d: D I, Fig. S3e**). Interestingly, analyzing the druggable genome list by excluding FDA approved drug targets showed underrepresentation of indispensable nodes (**Fig. 2d: D II** and **Fig. S3e**). This suggests a potential application of our analysis to redefine the druggable genome based on the network controllability.

All the above analysis of disease mutations, viruses and drugs consistently showed that indispensable nodes are preferred targets. We also analyzed how often indispensable nodes act as driver nodes by using a recently developed approach to identify the role of each node as drivers in the minimum driver node sets (MDSs) (16). We found that 378 nodes appears in all MDSs, i.e. they play roles in all the control configurations; 3,330 nodes are in some but not all MDSs, i.e. they play roles in some control configurations but the network can still be controlled without directly controlling them; and 2,631 nodes do not belong to any MDS, i.e. they play no roles in control (**Table S1**)(16). Interestingly, we found that indispensable nodes are never driver nodes in any MDS (**Fig. S3g** and **Table S1**). This fact can actually be rigorously proven (see Material and Methods). Moreover, perturbing indispensable nodes increases the number of driver nodes to control, suggesting that, from a controllability perspective, these nodes are fragile points in the network.

We further analyzed indispensable nodes in specific signaling pathways such as Receptor Tyrosine Kinase (RTK) signaling pathways, which are commonly perturbed in cancer(34). Strikingly, 67 out of 170 RTK pathway members are indispensable nodes (*p*-value < 0.0001), including 51 indispensable nodes targeted by disease mutations, viruses or drugs (**Fig. 2d** and **Table S2**). Further, we identified 21 indispensable nodes from different signaling pathways that are shared targets of cancer mutations, viruses as well as drugs (**Fig. 2e** and **Table S2**).

**Robustness of indispensable node classification**

The false-positive and false-negative interactions are major concerns in PPI networks, especially the false-negatives as the current networks are vastly incomplete(35). Hence, we systematically analyzed the robustness of node classification with respect to adding or removing interactions. Specifically, we analyzed the indispensable node classification as a function of removing edges (or network-filtering). The network-filtering is achieved by using a confidence score assigned to edge directions, where the most stringent filtering resulted in smaller high-confidence directed networks (20,151 edges and 5,317 nodes). We analyzed the controllability of filtered networks and compared it to the original network. The results show that 90% of the indispensable nodes in the stringent filtered network are indispensable in the original network (**Fig. 3a, Fig. S4a-b** and **Table S3**), suggesting that the indispensable node classification is robust with respect to adding or removing edges in the network.

Next, we analyzed the controllability of networks with perturbations, e.g. edge-rewiring or edge-direction flipping. In case of random rewiring, up to 100% of the edges are rewired (node degrees are preserved) and in case of direction-flipped networks, up to 100% of the edge directions are reversed. We observed that up to 50% of indispensable nodes in the rewired or direction-flipped network do not agree with the original annotation, showing that indispensability is highly sensitive to the connectivity pattern and edge direction (**Fig. 3b, Fig. S4c-f** and **Table S3**). Comparing indispensable nodes of real network to that of rewired (100% rewiring) and flipped (40% flipping) network revealed two subtypes (type-I and type-II) of indispensable nodes (**Fig. 3c** and **Table S3**). If a node's indispensability is robust to rewiring or flipping then we call it a type-I node; if its indispensability is sensitive to rewiring or flipping then we refer to them as type-II nodes. We found that 57% of indispensable nodes are type-I nodes and 43% are type-II.

Degree distribution of the subtypes shows that type-I nodes tend to be hubs, whereas the average degree of type-II nodes is similar to the average degree of the network (**Fig. 3d**). Indeed, type-II nodes cannot be distinguished from the rest of the nodes based on any other network properties analyzed (**Fig. S4g**). Further, type-I nodes show literature and annotation bias compared to type-II nodes (**Fig. 3e-f**). With respect to diseases, both node types show similar enrichment for cancer genes and other human diseases (**Fig. 3g**). In contrast to type-I nodes that tend to be hubs and well-studied genes, type-II nodes are poorly studied, show no special network feature except indispensability. This suggests that control theory brings orthogonal information to traditional network analysis.

**Applying network controllability analysis to mine cancer genomic data**

Our finding that indispensable nodes (both type-I and type-II) are more likely to correspond to cancer genes prompted us to systematically survey the perturbation of those genes in cancer. We analyzed data from 1,547 patients obtained from The Cancer Genome Atlas (TCGA; http://cancergenome.nih.gov/) and cBioPortal for Cancer Genomics(36), representing nine different cancer types (**Table S4**). Specifically, we analyzed the amplification or deletion of type-II indispensable nodes in nine cancer types. Note that the copy number alteration (CNA) data is normalized to the expression levels to identify the amplification or deletion that results in expression level changes (see methods). We ranked all genes based on the number of patients where the gene is amplified or deleted, and selected the top 1% as frequently amplified/deleted genes. 56 type-II genes were identified as part of the top 1% of deleted/amplified genes in nine cancer types (**Fig. 4a, Table S4**). Strikingly, 10 out of 56 type-II genes are known cancer genes, an overlap that is highly significant ($p$-value = 0.00002) (**Fig. 4b** and **Fig. S5a**). Interestingly, the frequency of deletion and amplification of type-II indispensable nodes are not significantly enriched compared to random sets, an observation that was similar to cancer gene census gene list (see **Table S4**). Further, we compared the type-II genes with results from a cell proliferation screen(37) that identified a subset of genes that regulate cell proliferation ("GO" genes induce the proliferation and "STOP" genes suppress the proliferation). 17 out of 56 genes represent regulators of cell proliferation (11 GO genes, 8 STOP genes and two genes part of both GO and STOP genes) (**Fig. 4c** and **Fig. S5b-c**). The overlap between type-II genes and GO genes are statistically significant (*p*-value = 0.0003). Out of 56 genes, 10 genes are frequently perturbed in multiple cancer types (e.g. proteasome 26S subunit, non-ATPase, 4 (PSMD4) in four different cancers) and all of them show similar deletion or amplification profile (e.g. PSMD4 amplified in all four cancers) (**Fig. 4d**). Almost half of the genes (23 genes) are poorly studied with less than 50 associated PubMed records, for instance small G protein signaling modulator 2(SGSM2) is associated with only 8 PubMed records (**Fig. 4d**). These contextual evidences along with the indispensability suggest that these 46 novel type-II nodes could be potential cancer genes.

**Database of directed PPI network with predicted controllability**

We created the DirectedPPI database (http://www.flyrnai.org/DirectedPPI/) to navigate the directed human PPI network with predicted controllability. Users can enter a gene or upload a list of genes and our tool generates a network with directed edges connecting the input list. Our tool also accepts gene list with values e.g. mutation frequency, p-values from GWAS, or expression changes. Three different node types (indispensable, neutral and dispensable) are distinguished with node shape and color and for these nodes all the properties analyzed in this manuscript are displayed. This tool will be useful to analyze disease datasets and other high throughput datasets to identify indispensable nodes and their interconnections.

**Discussion**

Studying the controllability of a complex biological network is rather difficult, due to the fact that we typically do not know the true functional form of the underlying dynamics. However, most biological systems operate near homeostasis, so local properties are indeed what we want to ascertain. Here we showed that application of linear control tools to study the local structural controllability of inherent non-linear biological networks provides meaningful predictions. Furthermore, we demonstrated that local controllability tools help identifies known human diseases genes and this can be used to identify novel disease genes and drug targets.

Our analysis of directed human PPI network identifies 36% of the nodes as driver nodes, which is similar to what has been observed in metabolic networks (~30%)(14). The node classification based on network controllability shows distinct biological properties in the context of essentiality, conservation and regulation. Specifically, we found that indispensable nodes are well conserved, highly regulated at the level of translational and posttranslational modifications and important for the transition between healthy and disease states. Interestingly, this enrichment pattern is partially shared by the nodes that are located in strategically important positions in the network(38). Furthermore, identification of the indispensable nodes as primary targets of diseases causing mutations, viruses and drugs revealed a potential application of this framework to identify novel disease genes and potential drug targets.

Interestingly, disease causing mutations, viruses and drugs, target fragile points (indispensable nodes) that determine the number of driver nodes rather than the driver nodes themselves, suggesting that network controllability is crucial in transitioning between healthy and disease states. Although network topology based properties such as hubs and modules are commonly used to identify disease genes(39-42), the controllability perspective provides a complementary network analysis framework for network medicine. Especially, type-II nodes that are not distinguishable from existing network properties and without publications bias were still identified by controllability framework as nodes of special interest. We envision that in the future, improving the quality and the completeness of interactome maps, and integrating dynamics of network components would hugely impact our understanding of biological networks both in the context of biological function and human disease.

## Materials and Methods

### Directed human PPI network

The directed human PPI network was compiled from our previous study(13). Briefly, a Naïve Bayesian classifier was applied to predict potential direction of signal flow between the *i*-th and *j*-th interacting proteins $p_i$ and $p_j$ as $p_i \rightarrow p_j$, $p_j \rightarrow p_i$, or both. The classifier uses features derived from the shortest PPI paths between membrane receptors and transcription factors and assigns confidence for each predicted edge directions ranging from 0.5 to 1. The weighted and directed edges are then encoded in an $N \times N$ matrix, $A$ denoted as the weighted adjacency matrix of the directed graph for the PPI network. The element of $A$ in the *i*-th row and *j*-th column is denoted as $a_{ij}$ and is defined as follows, $a_{ij}$ is in the range [0.5,1] if there is signal flow from protein $p_j$ to $p_i$ otherwise $a_{ij} = 0$.

### Controllability Analysis and Node Classification

Recently, we developed a mathematical framework and analytical tools to identify a minimum driver node set (MDS), with size denoted as $N_D$, whose control is sufficient to ensure the structural controllability of linear dynamics (14) and local structural controllability for nonlinear dynamics (see Supplement Text) on any directed weighted network. This is achieved by mapping the structural controllability problem in control theory to the maximum matching problem in graph theory, which can be solved in polynomial time(15). Here, an edge subset $M$ in a directed network or digraph is called a *matching* if no two edges in $M$ share a common starting node or a common ending node. A node is *matched* if it is an ending node of an edge in the matching. Otherwise, it is *unmatched*. A matching of maximum cardinality is called a *maximum matching*. (In general there could be many different maximum matchings for a given digraph.) We proved that the unmatched nodes that correspond to any maximum matching can be chosen as driver nodes to control the whole network. Identifying a minimum set of driver nodes is equivalent to choosing an input matrix (often denoted as *B*) with minimum number of columns (See Supplement Text and (14) for more details). The detailed construction of the input matrix $B$ is not necessary for the identification of driver nodes. This is only mentioned to connect the notion of a driver node to the theoretical discussions in Supplement Text.

After a node is removed, denote the minimum number of driver nodes of the damaged network as $N_D'$. In this work we classified nodes into three categories: 1) A node is *indispensable* if in its absence we have to control more driver nodes, i.e. $N_D' > N_D$. For example, remove one node in the middle of a directed path will cause the $N_D$ increase. 2) A node is *dispensable* if in its absence we have $N_D' < N_D$. For example, removal of one leaf node in a star will decrease $N_D$ by 1 3). A node is *neutral* if in its absence $N_D' = N_D$. For example, removal of the central hub in a star will not change $N_D$ at all.

Note that indispensable nodes are never driver nodes in any control configurations or MDSs. This can be proven by contradiction. Assume a driver node $i$ is indispensable. According to the minimum input theorem (9), driver nodes are just unmatched nodes with respect to a particular maximum matching. There are two cases: (1) node $i$ has no downstream neighbors (i.e., $k_{out}(i)=0$), then in its absence $N_D' = N_D-1$. (2) node $i$ has at least one downstream neighbors (i.e., $k_{out}(i)>0$) and we assume in the maximum matching one of its neighbors (node $j$) is matched by node $i$. Then in the absence of the driver node $i$, node $j$ will become unmatched (i.e., a new driver node), rendering $N_D' = N_D$. In both cases, we don't have $N_D' > N_D$, which is in contrast to the definition of indispensable nodes. Hence driver nodes cannot be indispensible.

**Datasets used for enrichment analysis**
All the datasets used for the enrichment analysis in this study were listed in Table S2. This includes the source of the data, reference, number of proteins compiled and overlap with human directed PPI network. The datasets were downloaded from respective databases or publications as mentioned in Table S2. The gene or protein ids from various resources were mapped to Entrez gene IDs. All compiled datasets are available as an integrated table that shows the nodes and the overlap with respective datasets (**Table S1**).

### Enrichment analysis

To estimate the significance of overlap between a given node type $S$ and given dataset D, we compute an enrichment Z score as

$$Z\ score = \frac{(S_D - mean\ of\ R_D)}{Standard\ deviation\ of\ R_D}$$

Where $S_D$ is number of proteins from dataset $D$ overlapping with node type $S$, $R_D$ is number of proteins from dataset $D$ overlapping with random set of proteins of same size as $N$. Mean and standard deviation of $R_D$ is computed from 1,000 simulations of random sets. Note that the entire network with 6,339 proteins is used as the background for random sampling. In addition to the Z score, we also computed p value (two-tailed) by comparing the $S_D$ with $R_D$ distribution (modeled as Gaussian distribution). In case of degree or literature controlled random sets, the random sets are sampled such that the average degree or average PubMed records of random sets matches the average of node type $S$.

### Random networks

To compare the real network with its randomized counterparts, we performed two types of randomization: 1) edge-rewiring: we randomly choose a $p$ fraction of edges to rewire, using the degree-preserving random rewiring algorithm(43); 2) edge-flipping: we randomly choose a $p$ fraction of edges to flip their directions. We tune $p$ from 0 up to 1, resulting in a series of randomized networks.

### Analysis of cancer genomic datasets

Copy number alteration data for nine cancer types were downloaded from the cBioPortal for Cancer Genomics (version corresponds to April 2013, http://www.cbioportal.org/public-portal/). Using GISTIC algorithm(44) the cBioPortal provides putative values of copy number alterations for each cancer patient. The GISTIC score -2, -1, 0, 1, 2, corresponds to deep loss (possibly a homozygous deletion), single copy loss (heterozygous deletion), diploid, low level gain and high amplification respectively. The gene expression data for each cancer type were downloaded from The Cancer Genome Atlas (TCGA, version corresponds to April 2013, https://tcga-data.nci.nih.gov/tcga/). The tumor-matched datasets (for each participant have been analyzed and compared with normal tissue on the CNA and gene expression level) were used in the analysis. TGCA data Level-3 (expression calls for genes, per sample) was used in our study. The TCGA data were downloaded by using TCGA web interface with filters set as "Data Type: Expression-Genes"; "Data Level: Level 3"; "Tumor/Normal: Tumor-matched".

Next, we filtered for patients with both CNA and expression data available (details are available in **Table S4**). We computed a Z-score for each gene in a patient to identify whether the amplification or deletion results in expression change for the corresponding gene. Briefly, for each gene the diploid mean and standard deviation of expression values were calculated using the data from patients without any copy number alteration (Gistic score "0", diploid). Using the diploid mean and standard deviation, we computed z-score for each gene in a given patient. A gene is defined as amplified if the GISTIC score is ≥ 1 and the z-score ≥ 1.5, and deleted if the GISTIC score is ≤ -1 and the z-score ≤ -1.5. All the data preprocessing and normalization were performed using Perl and Java scripts developed in house.

**Acknowledgments**: We thank M.W. Kirschner, S.E. Mohr, I.T. Flockhart and S. Rajagopal for helpful suggestions. Funding: This work was supported by National Institutes of Health (NIH) grants P01-CA120964, R01-GM067761, R01DK088718, P50-HG004233-CEGS, MapGen grant (1U01HL108630-01), 5PO1-HL083069-5, 1-RC HL10154301-2, and the John Templeton Foundation (award number 51977). N.P. is supported by Howard Hughes Medical Institute. **Author contributions**: A.V., Y.Y.L., N.P. and A.L.B conceived the project. All authors contributed to the manuscript preparation and A.V. leads the writing of the manuscript. A.V. performed the enrichment analysis. Y.Y.L. developed and performed structural controllability analysis and node classification. T.E.G. contributed the local controllability analysis. B.Y., C.R and A.V. developed the DirectedPPI database. A.V and B.Y. analyzed the cancer genomics datasets. H.J.L., Y.H., Y.K., and A.S. contributed analysis and datasets. **Competing interests:** The authors declare that they have no competing interests.

**Figure legend**

**Fig. 1**. Characterizing the controllability of human directed PPI network. **a**) Schematic representation of the node classification using controllability framework. **b**) Identification of indispensable, neutral and dispensable nodes in human directed PPI network. **c**) In-degree distribution and average in-degree for three different node types. **d**) Out-degree distribution and average out-degree for three different node types. **e**) Distinct enrichment profiles of indispensable, neutral and dispensable nodes in the context of essential genes, evolutionary conservation, cell signaling, protein abundance and post-translational modifications.

**Fig. 2**. Characterizing network controllability in transition from healthy to disease state. **a**) Bar graph showing the enrichment results (Z scores) of cancer genes compared to the random sets (Cancer I = cancer gene census) and the random sets controlled for literature (PubMed) or Degree (Degree) bias. In case of degree or literature controlled random sets, the random sets are sampled such that the average degree or average PubMed records of random sets matches the average of node type N. **b**) Results from enrichment analysis of dataset corresponding to extended list of cancer genes (Cancer II), other human diseases (OMIM) and genome-wide association studies (GWAS). **c**) Results from enrichment analysis of the targets of HIV identified using RNAi screens (RNAi) and PPI networks (PPIs), and targets of other human virus (208 viruses). **d**) Enrichment results from targets of FDA approved drugs and druggable genome (DI = druggable genome; DII = druggable genome excluding FDA approved targets). **e**) Members of receptor tyrosine signaling pathways that are predicted as indispensable nodes and targeted by cancer mutations, OMIM disease, viruses or FDA approved drugs. RTK pathway members as defined by SignaLink database(45). **f**) Indispensable nodes that are targeted by all three inputs (cancer mutation, viruses and drugs). The labels of FDA drug nodes correspond to DrugBank IDs. The network was generated using Cytoscape(46).

**Fig. 3**. Perturbation of network connectivity reveals two sub types of indispensable nodes (Type-I and Type-II). **a**) Plot showing the fraction of indispensable nodes in filtered networks that overlaps with real network. The network filtering achieved using edge confidence score. **b**) Fraction of indispensable nodes in rewired or direction-flipped overlap with real network. **c**) Identification of type-I and type-II indispensable nodes. The average node degree (**d**), PubMed record association (**e**) and Gene Ontology term annotations (**f**) for type-I and type-II indispensable nodes. **g**) Enrichment of type-I and type-II indispensable nodes as cancer genes and OMIM disease genes.

**Fig. 4**. Applying network controllability to mine cancer genomic data. **a**) Type-II genes frequently amplified or deleted in cancer patients (part of top 1% genes). The bar plot shows number of type-II genes deleted in Brain Lower Grade Glioma (LGG), Kidney Renal Clear Cell Carcinoma (KIRC), Lung adenocarcinoma (LUAD), Lung squamous cell carcinoma (LUSC), Ovarian serous cystadenocarcinoma (OV), Uterine Corpus Endometrial Carcinoma (UCEC), Breast invasive carcinoma (BRCA), Colon Adenocarcinoma (COAD), Glioblastoma Multiforme (GBM) cancers. **b**) Overlap between frequently deleted/amplified type-II genes and known cancer genes. **c**) Overlap between frequently deleted/amplified type-II genes and regulators of cell proliferation (STOP genes reduces cell proliferation and GO genes increases cell proliferation). The *p* values show the significance of overlap calculated based on 1000 random sets. **d**) Network representation of 56 type-II genes frequently deleted (red edge) or amplified (blue edge) in nine different cancer types. The node size corresponds to the number of PubMed records associated with the gene.

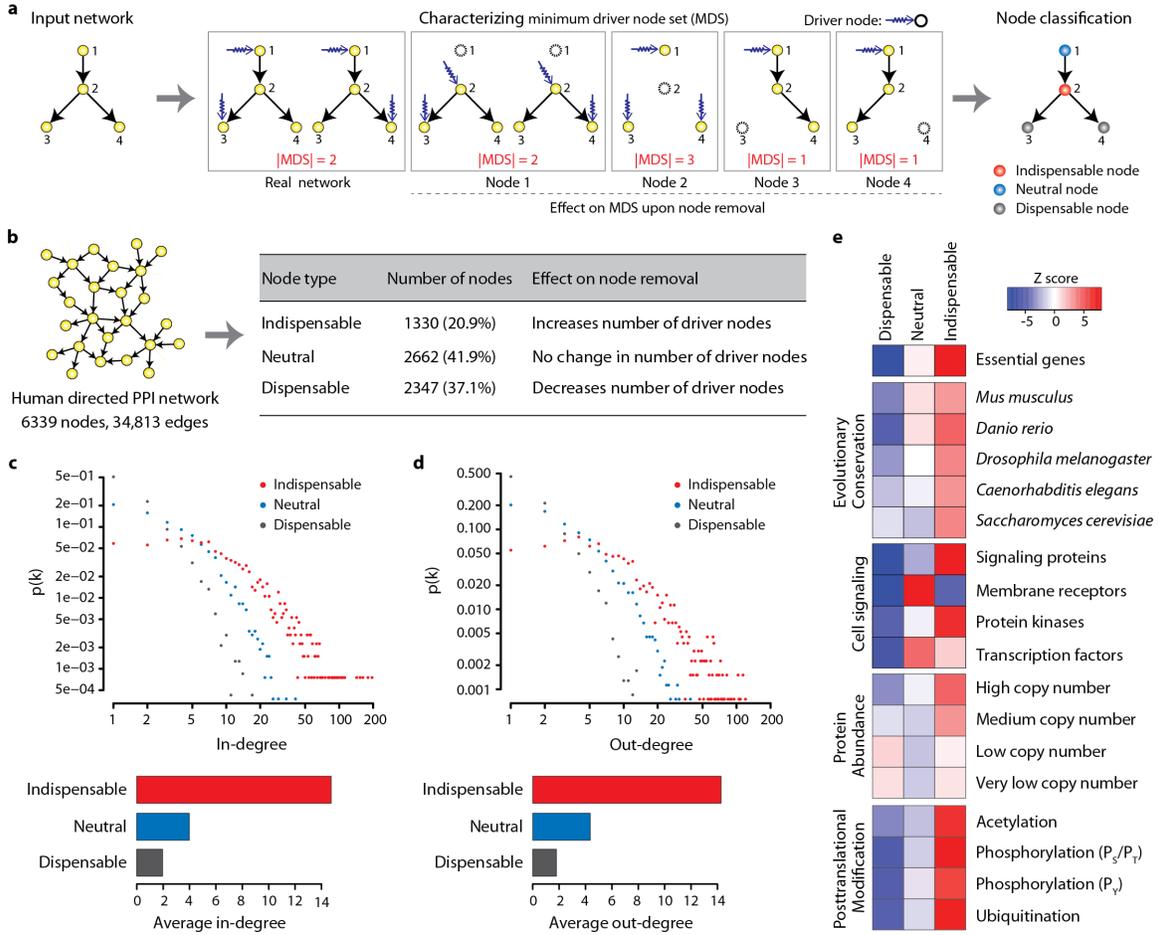

Fig.1

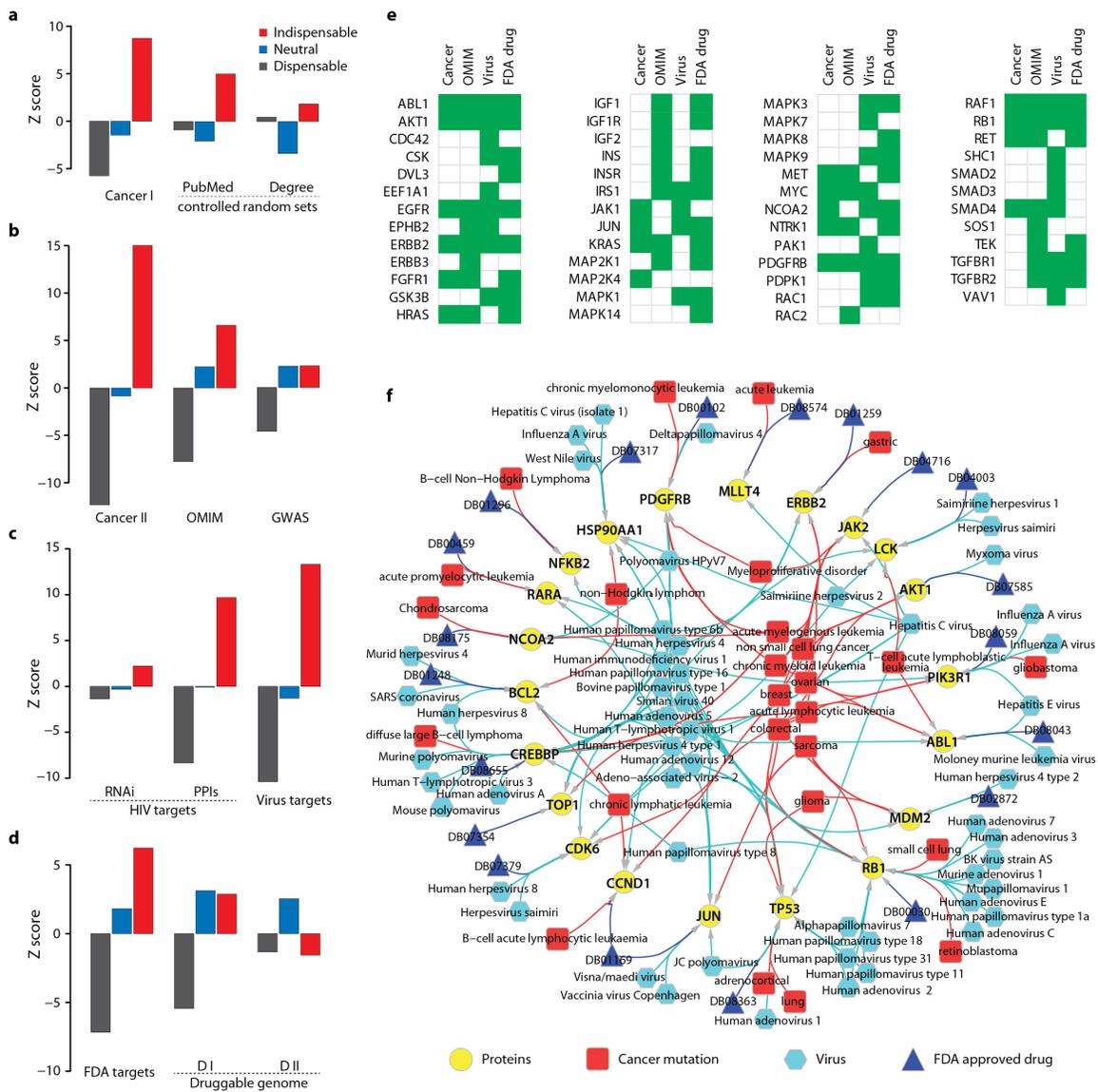

Fig.2

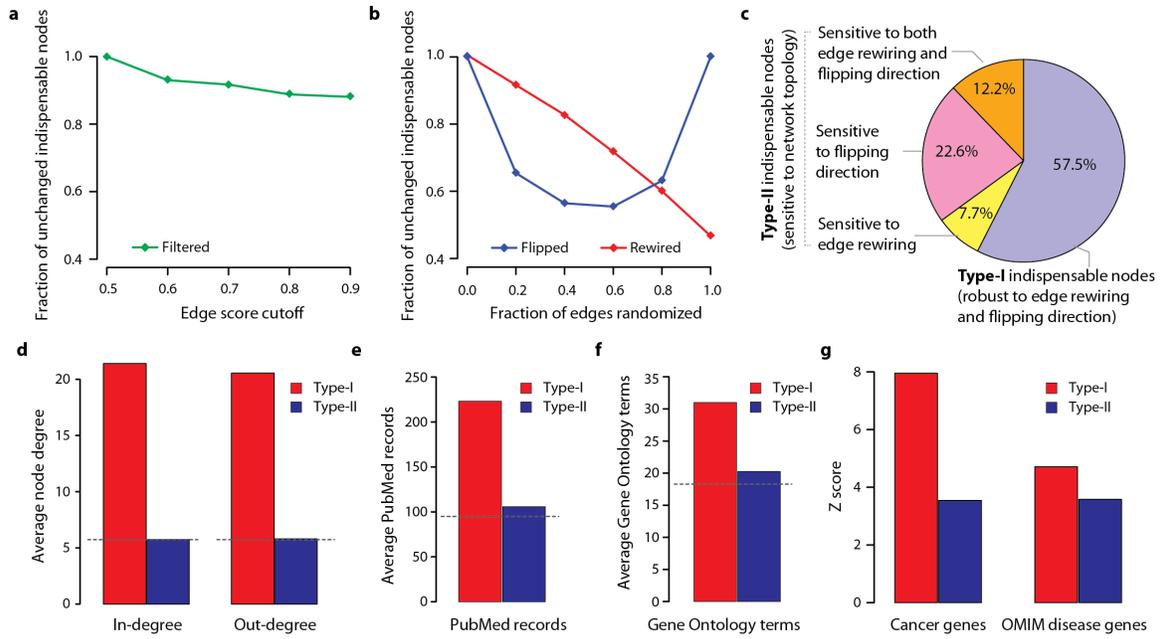

Fig.3

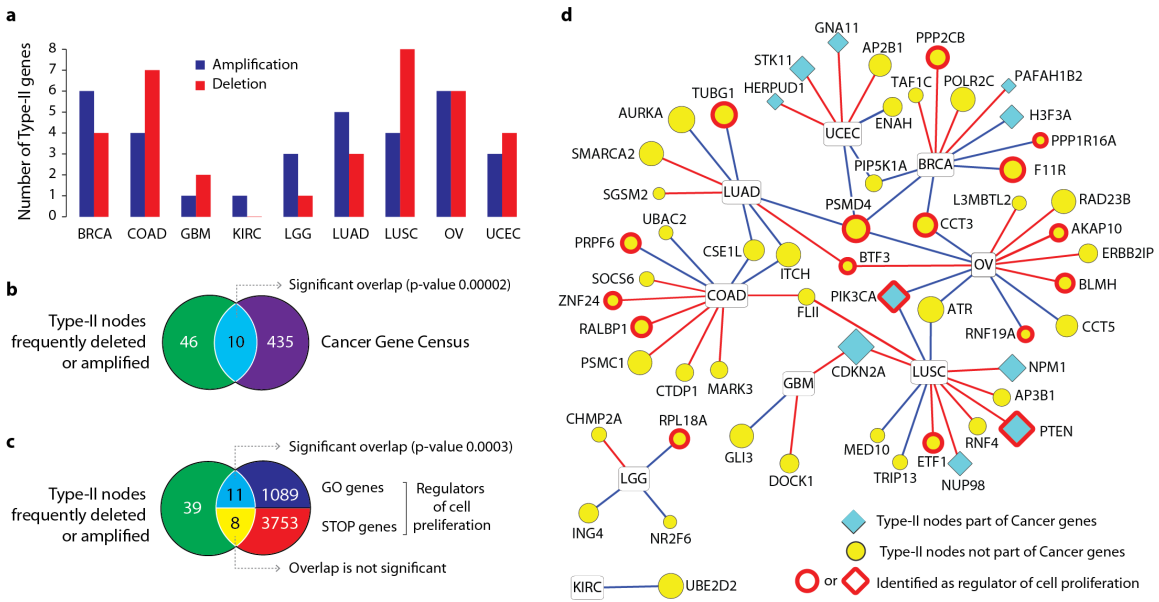

Fig.4